\documentclass[letterpaper]{article} 
\usepackage[draft]{aaai25}  
\usepackage{times}  
\usepackage{helvet}  
\usepackage{courier}  
\usepackage[hyphens]{url}  
\usepackage{graphicx} 
\usepackage{natbib}  
\usepackage{caption} 
\usepackage{algorithm}
\usepackage{algorithmic}

\usepackage{newfloat}
\usepackage{listings}
\DeclareCaptionStyle{ruled}{labelfont=normalfont,labelsep=colon,strut=off} 
\floatstyle{ruled}
\newfloat{listing}{tb}{lst}{}
\floatname{listing}{Listing}
\usepackage{multirow}

\usepackage{graphicx}
\usepackage{dblfloatfix}
\usepackage{caption}
\usepackage{subcaption}

\usepackage{amsthm}

\makeatletter
\def\thm@space@setup{\thm@preskip=0pt
\thm@postskip=0pt}
\makeatother
\newtheoremstyle{newstyle}      
{5pt} 
{} 
{\itshape} 
{} 
{\bfseries} 
{.} 
{ } 
{} 

\theoremstyle{newstyle}
\newtheorem{theorem2}{Research Question}



\usepackage{censor}
\usepackage{pifont}
\newcommand{\cmark}{\ding{51}}%
\newcommand{\xmark}{\ding{55}}%

\title{Do Large Language Models Learn Human-Like Strategic Preferences?}
\author {
    Jesse Roberts\textsuperscript{\rm 1,2},
    Kyle Moore\textsuperscript{\rm 2},
    Doug Fisher \textsuperscript{\rm 2},
}
\affiliations {
    \textsuperscript{\rm 1}Tennessee Technological University\\
    \textsuperscript{\rm 2}Vanderbilt University\\
    Jesse.TN.Roberts@gmail.com, Kyle.A.Moore@Vanderbilt.edu, Douglas.H.Fisher@Vanderbilt.edu
}

\usepackage{bibentry}

\begin{document}

\maketitle

\begin{abstract}
In this paper, we evaluate whether LLMs learn to make human-like preference judgements in strategic scenarios as compared with known empirical results. Solar and Mistral are shown to exhibit stable value-based preference consistent with humans and exhibit human-like preference for cooperation in the prisoner's dilemma (including stake-size effect) and traveler's dilemma (including penalty-size effect). We establish a relationship between model size, value-based preference, and superficiality. Finally, results here show that models tending to be less brittle have relied on sliding window attention suggesting a potential link. Additionally, we contribute a novel method for constructing preference relations from arbitrary LLMs and support for a hypothesis regarding human behavior in the traveler's dilemma.
\end{abstract}

%

\section{Introduction}

Transformer-based large language models (LLMs) have famously achieved state of the art performance on many tasks since their introduction by \citet{vaswani2017attention}. While the analysis of LLMs typically focuses on benchmark tasks like \cite{srivastava2022beyond}, MMLU \cite{hendrycks2020measuring}, and Agieval \cite{zhong2023agieval}. On the other hand, theoretical analysis of their computational abilities \cite{roberts2024powerful,bhattamishra2020computational,perez2019turing} and empirical investigations of their cognitive behaviors \cite{misra2021language,trott2023large,roberts2024using,binz2023using,ullman2023large,suri2023large} are less common. However, these latter analyses are of utmost importance in many human-adjacent cooperative applications. 

\subsubsection{Motivation} Consider a human carrying a heavy box who asks a collaborator for help. The individual asking for help implicitly relies upon the collaborator's possession of a compatible set of preferences over the possible strategies. Based on the request and visual input alone, the collaborator is expected to quickly choose and apply their most preferred strategic mixture of vertical and horizontal force. Otherwise, the originator of the request would need to provide more detailed and precise instructions to ensure appropriate action.

In contrast, a robot asked to help with a box is currently incapable of selecting from possible strategies unless imbued with a precise value function over the strategies or trained through reinforcement learning. We aim to apply LLMs to support this sort of natural language human-robot interaction (HRI) in future work. However, for natural language human-robot collaboration to be effective, a supporting LLM must have strategic preferences sufficiently similar to that of a human to permit effectual communication. 

Furthermore, applications like HRI require stable LLM behavior under variations to avoid potentially dangerous strategic variations due to slight contextual irregularities. This point is timely as recent LLM cognitive behavioral studies have been shown to not replicate under small variations \cite{ullman2023large}. We apply PopulationLM \cite{roberts2024using} to ensure empirical results are robust to systematic variations.

\subsubsection{Aims} In service of HRI LLM applications, this paper aims to understand if any current open-source language models exhibit stable, human-like strategic preferences. We choose empirical human behaviors from the field of game theory as the point of comparison and focus on open-source models to support reproducibility.

We begin by evaluating a large body of LLMs and identifying those that tend to have value-based preferences (VBP). We then engage the identified models in high and low stakes prisoner's dilemmas (PD) followed by high and low penalty traveler's dilemmas (TD) to characterize their similarity to nuanced human strategic preference for cooperation.

\subsubsection{Contributions} Our findings demonstrate that:

\begin{enumerate}
    \item Some LLMs acquire stable human-like strategic preferences. Specifically, we identify Solar \cite{kim2023solar} and Mistral \cite{jiang2023mistral} as potential models appropriate for HRI-related tasks.
    
    \item Small models tend to prefer strategies based on superficial heuristics, while larger models tend to have VBP.

    \item Most large models are brittle under variations, which we hypothesize may be related to the attention architecture.

    \item Models with stable VBP tend to have nuanced human-like strategic preference for cooperation.

    \item Deviation from the Nash equilibrium in the TD stems from penalty dependent uncertainty regarding weakly dominated strategies, which provides \textit{in silico} evidence for the analogous explanation in humans.
    
\end{enumerate}

Additionally, we propose a method for constructing Pythagorean preference relations from an LLM population.

%
%

\section{Related Work}

\begin{figure*}[h]
\centering
\includegraphics[width=0.9\linewidth]{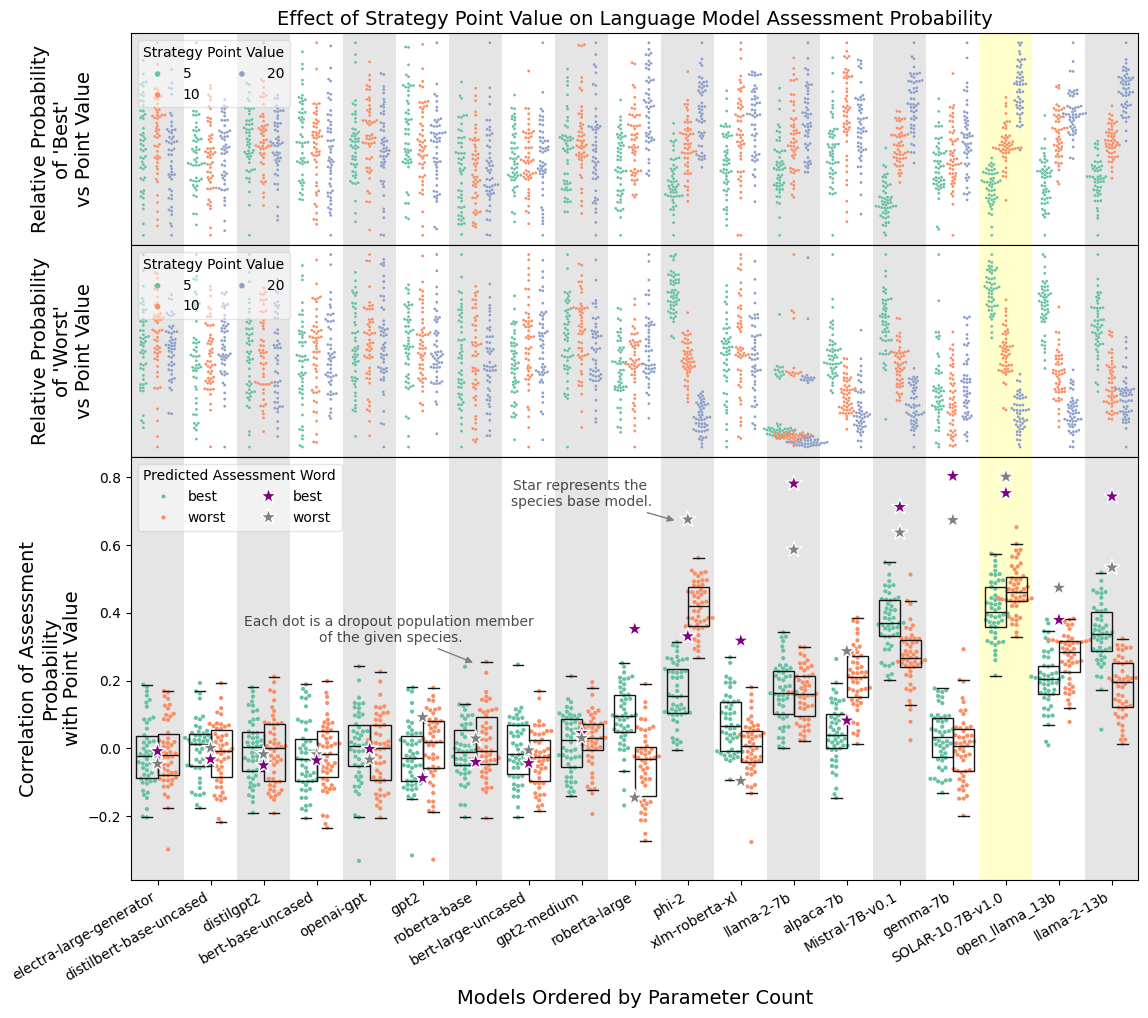} 
\caption[Summary of LLM Value-Based Preference Experiments]{Top: Population member probabilities for ``Best'' evaluation of strategies. Middle: Population member probabilities for ``Worst'' evaluation of strategies. Bottom: Spearman's $\rho$ for value-preference correlation and negated anti-correlation.}
\label{fig:value}
\end{figure*}


Several authors have explored LLM behavior in games, providing confidence that some LLMs may learn strategic preferences from human language data.

\citet{akata2023playing} engaged GPT-3.5 and GPT-4 \cite{OpenAI2023GPT4TR} in iterated games, including an iterated prisoner's dilemma. They found that both models tended to be punishing in response to betrayal, though they cooperated prior to betrayal. Interestingly, the models would not reciprocate cooperation after a betrayal, regardless of how many times an opponent cooperated subsequently.

\citet{fan2024can} evaluated GPT-3.5 and GPT-4's ability to act consistently with a prompted preference, refine belief, and take optimal actions in various games. Their work aimed to assess the potential integration of GPT-4 into games for social science research. Results suggest that GPT-4 fails to appropriately update and maintain beliefs necessary to choose optimal strategies, making it unsuitable for integration into social science experiments. However, GPT-4 is more common capable of choosing optimal strategies in typical scenarios.

Our work differs significantly from existing literature in terms of aims and methods. We specifically consider the nuanced cooperative strategic behavior of LLMs with systematic variations. Furthermore, we are the first to engage LLMs in a traveler's dilemma, where human behavior differs importantly from game-theoretic predictions. While existing work measures model preference using a cloze task, we use a counterfactual prompting method to measure model evaluation probability. Finally, we consider strategic capability in a wide array of open-source models and examine the role of model size in the presence of value-based preferences (VBP).

\begin{figure*}[h]
\centering

\begin{subfigure}[t]{.53\linewidth}
  \centering
  \includegraphics[width=0.98\linewidth, height=120px]{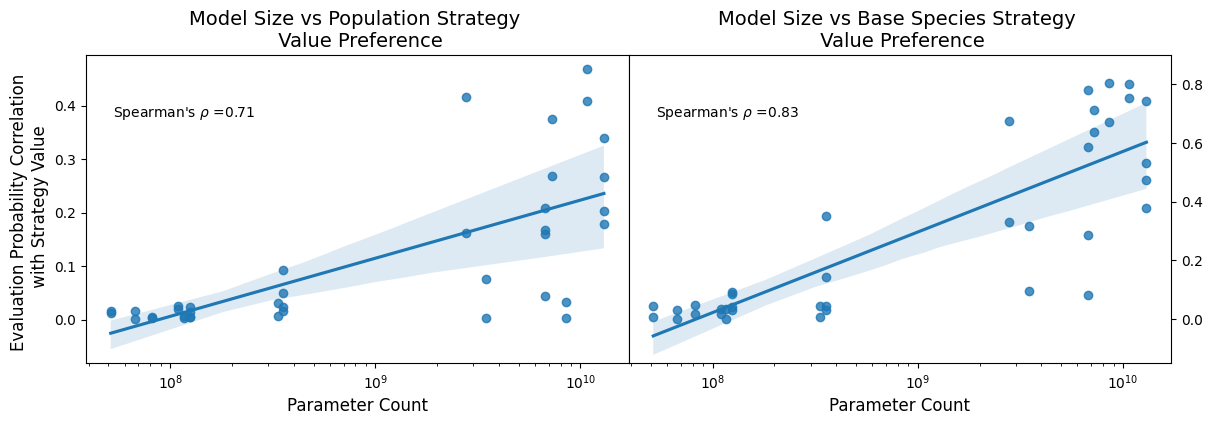}
\label{fig:modelSize}
\end{subfigure}%
\hfill
\begin{subfigure}[t]{.47\linewidth}
  \centering
  \includegraphics[width=0.98\linewidth, height=120px]{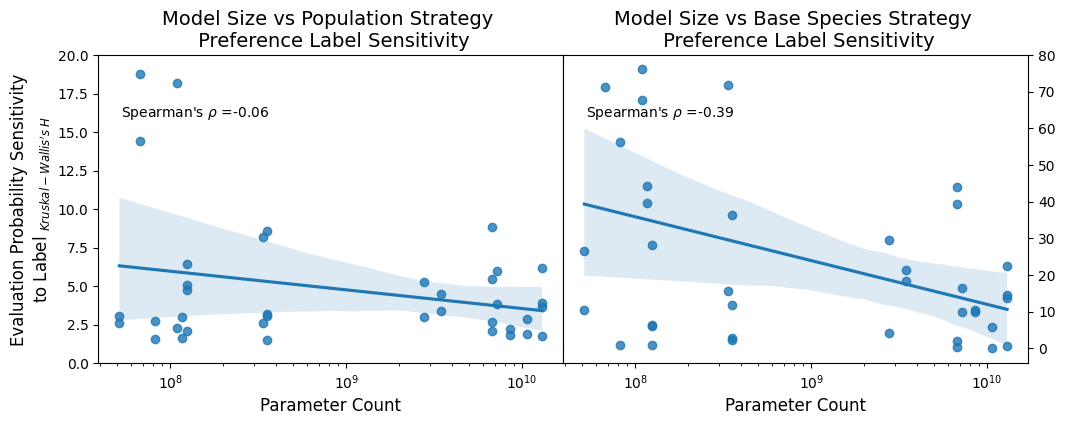}
  
\label{fig:modelSizeSuperficial}
\end{subfigure}%
\caption[Model Size Correlation with Value-Based Preference]{As models get larger they tend to have value-based strategy preferences and tend to be less sensitive to arbitrary labels. The strength of this relationship is largest in the base models suggesting the behavior is less typical in the population.}
\label{fig:groupModelSize}
\end{figure*}

\section{Do LLMs Prefer Strategies Based on Value?}

\subsubsection{The Aim} Previous research has demonstrated that GPT-3.5 and GPT-4 have preferences for higher-valued strategies in a dictator game \cite{fan2024can}. In this study, we extend that finding by evaluating how preferences relate to model size through the examination of value-based preference (VBP) in a larger body of models. Additionally, we apply systematic perturbation via PopulationLM to understand the stability of these preferences.

If systematic perturbation leads to brittle behavior, we consider a preference to be poorly supported. Poorly supported preferences in a model may not be sufficiently reliable to support human-adjacent NLP tasks. This leads us to formulate RQ\ref{RQ:VBP}.

\begin{theorem2}
Given a set of strategies each with a clearly specified value, do LLMs have stable value-based preferences, and how do these preferences relate to model size?
\label{RQ:VBP}
\end{theorem2}

\subsection{Methodology}

To evaluate RQ \ref{RQ:VBP}, we create a prompt that defines 3 strategies labeled A1, A2, and A3. Each strategy is ascribed a value 5, 10, or 20 points with each value being assigned once in the prompt context $c$. The model then provides the probability for all in-vocabulary completions. However, we consider only the probability of a constant evaluation word. This is repeated for each strategy option, changing only the strategy $s$. This measures the probability of the evaluation word given the strategy, $p(e_{word}|c,s)$ $\forall s\in \mathbf{S}$. We refer to this as \textit{counterfactual prompting}. The following is an example of such a prompt with \textit{A1} as the evaluated strategy. 

\textit{Option A1 gives 5 points. Option A2 gives 10 points. Option A3 gives 20 points. A1 is} \blackout{best.}

We hypothesize that the preference, as measured by the probability of the evaluation word, will tend to be correlated with the assigned value. Based on \textit{Applied Statistics for the Behavioral Sciences} \cite{hinkle2003applied}, if the correlation is 0.3 or higher, a significant correlation is present, and the LLM is considered capable of VBP. To control for alternative hypotheses of preference based on label ordering or preference for a label absent of value, we generate a prompt for every permutation of the order of labels and the assigned value, resulting in 36 unique prompts. We then test the LLM preference for each strategy for each prompt permutation. We test the LLM preference for each strategy for each prompt permutation, yielding 108 individual experiments per model population member (N=50).

Furthermore, we investigate if models with VBP are self-consistent across evaluation words of differing sentiment. We perform the experiment first with a positive sentiment evaluation word (best) and then with a negative sentiment evaluation word (worst). A model is considered to have VBP and be \textit{self-consistent} if the positive sentiment probability is correlated with strategy value and the negative sentiment probability is anti-correlated with strategy value.

Given the targeted HRI application domain, the effect of variation on model preference is crucial. We use PopulationLM \cite{roberts2024using} to construct populations for each model species tested. Models that differ on architecture, size, training data, or training task are considered different species. This approach uses Monte Carlo dropout to generate perturbed versions of the base model, approximating a Gaussian random process \cite{gal2016dropout}.  Intuitively, model behaviors constituted in a small number of neurons, referred to as poorly supported, are more likely to be ablated in the perturbed population than those that are more distributed. If the base model of a given species has VBP, but the derived population does not, we say the model is \textit{brittle} since variation tends to erode the behavior of interest.

Finally, to understand how model size relates to VBP and the tendency to prefer strategies based on more superficial criteria, we conduct the described set of experiments on 19 different model species with sizes varying from $<10^{8}$ to $>10^{10}$ parameters.

\subsection{Results: Value-Based Preference}

In answer to RQ \ref{RQ:VBP}, we find that a surprisingly small number of models have VBP. In figure \ref{fig:value}, the correlation of the evaluation probability and strategy point value for each of the population members (dots) as well as the species base model (stars) are shown in the bottom row. Those that do have high base model correlation like Solar. Table \ref{tab:VBP} gives a summary of results for models with VBP.

\vskip-0.2em

\begin{table}[h]
\caption[VBP]{SOLAR \& Mistral have stable, self-consistent VBP}
\resizebox{0.95\linewidth}{!}{%
\begin{tabular}{ lcccc }
\hline
Model & Paper & VBP & Self-Consistent & Stable \\
\hline
Solar &  \citeauthor{kim2023solar}   & \cmark & \cmark             & \xmark      \\
Mistral & \citeauthor{jiang2023mistral} & \cmark & \cmark             & \xmark      \\
Gemma & \citeauthor{team2024gemma} & \cmark & \cmark             & \cmark     \\
Llama-2 & \citeauthor{touvron2023llama} & \cmark & \cmark             & \cmark     \\
Phi-2 & \citeauthor{javaheripi2023phi}  & \cmark & \xmark             & \xmark      \\
\hline
\end{tabular}%
}
\label{tab:VBP}
\end{table}

\vskip-0.2em

The brittleness of Gemma and Llama-2 models raises concerns about their reliability in real-world applications, particularly in human-robot interaction (HRI) scenarios where consistent value-based decision-making is crucial. On the other hand, the stability of VBP in Solar and Mistral suggests that these models may be more suitable for HRI tasks.

\subsection{Effects of Model Size}

We investigate the effect of model size on the presence of VBP. Figure \ref{fig:groupModelSize} (left) shows a telling correlation between model size and the model's preference for higher-value strategies. This suggests that model size may be predictive of VBP. More precisely, we conclude that sufficient model size may be a necessary, though insufficient, condition for a model to learn VBP from human language data.

We further consider the effect of superficial information, like the label, on model preference. Figure \ref{fig:groupModelSize} (right) uses the non-parametric Kruskal-Wallis test to evaluate if the probabilities assigned to a strategy are independent of the label. The null hypothesis for this test expects the medians of the groups to be equal. The figure shows that preferences in smaller species base models tend to be sensitive to superficial information like labels. However, as model size increases, sensitivity to the label tends to decrease.

Interestingly, preference sensitivity to label appears to be much more correlated with model size in the base models ($\rho=0.39$, shown on the right of the figure) compared to the populations ($\rho=0.06$, shown on the left). This indicates that intra-species populations of language models may tend to be less sensitive to superficial information. In other words, the sections of the network that respond to superficial information tend to be ablated in much of the population.

\subsection{Unraveling the Robustness of Solar and Mistral}

Our experiments reveal that Solar and Mistral excel at making stable value-based preference (VBP) judgments, while Gemma and Llama-2 exhibit brittleness despite comparable VBP. This disparity raises the question: what sets Solar and Mistral apart?

To begin to answer this, we must examine the origins of these models. Mistral builds upon Llama-2, which was trained on 2 trillion tokens but had not reached saturation \cite{touvron2023llama}. Mistral's creators incorporated sliding window attention (SWA) \cite{beltagy2020longformer} into Llama-2's architecture and retrained the model from the pre-trained weights. SWA requires the model to channel information from tokens prior to the window through adjacent latent representations, resulting in substantial performance gains over Llama-2 7B and 13B \cite{jiang2023mistral}.

Solar, in turn, adopted Llama-2's architecture, increased the number of layers through depth upscaling \cite{kim2023solar}, and initialized its initial layers with Mistral's weights before additional training. Solar must therefore be considered to have been trained on more tokens than Mistral. While Solar doesn't employ SWA directly, it inherits the benefits of Mistral's SWA-learned weights.

Interestingly, Gemma exhibits VBP consistent with Solar but is more brittle than Llama-2, despite being trained on 4 times the number of tokens. This suggests that while training tokens and model size may improve VBP, they are insufficient for reducing brittleness.

We hypothesize that SWA may encourage a more distributed representation, leading to reduced brittleness. However, this \textbf{reasoning is not conclusive}. To resolve this, \textbf{future work} should focus on understanding how \textbf{SWA} impacts learned \textbf{representations} to develop more resilient language models.

\section{Do LLMs Have Human-Like Preference in the Prisoner's Dilemma?}

The models found to have stable VBP are further evaluated in comparison to human-like cooperative preferences in the prisoner's dilemma (PD). Those without self-consistent VBP are not expected to exhibit more sophisticated preferences and are therefore not included in additional experiments. 

\subsubsection{The Game} The PD is a well-known game in which two players each have two strategy options: betray or remain silent. The payoff for each player depends on the combination of their chosen strategies. Table \ref{tab:prisonersdilemma} shows the payoff matrices for various scenarios, with Player 1's strategy being the first in each ordered pair.

\begin{table}[h]
\caption[Prisoner's Dilemma Payoff Matrices]{Two Player Prisoner's Dilemma Payoff Matrices}
\resizebox{0.95\linewidth}{!}{%
\begin{tabular}{lccccccc}                                                                                  \cline{3-8} 
                                               & \multicolumn{1}{c|}{}       & \multicolumn{2}{c|}{AC Sharing}                               & \multicolumn{2}{c|}{Life Support Sharing}                 & \multicolumn{2}{c|}{Time in Jail}                       \\ \cline{3-8} 
                                               & \multicolumn{1}{c|}{}       & \multicolumn{1}{c|}{Silent} & \multicolumn{1}{c|}{Betray}     & \multicolumn{1}{c|}{Silent} & \multicolumn{1}{c|}{Betray} & \multicolumn{1}{c|}{Silent} & \multicolumn{1}{c|}{Betray} \\ \cline{2-8} 
\multicolumn{1}{c|}{\multirow{2}{*}{}} & \multicolumn{1}{c|}{Silent} & \textbf{Cool, Cool}                  & \multicolumn{1}{c|}{Cold, Hot}  & 4,4                         & \multicolumn{1}{c|}{0,10}   & 2,2                         & \multicolumn{1}{c|}{5,0}    \\ \cline{2-2}
\multicolumn{1}{c|}{}                          & \multicolumn{1}{c|}{Betray} & Cold, Hot                   & \multicolumn{1}{c|}{\textit{Warm, Warm}} & 10,0                        & \multicolumn{1}{c|}{\textit{\textbf{2,2}}}    & 0,5                         & \multicolumn{1}{c|}{\textit{\textbf{3,3}}}    \\ \cline{2-8} 
\end{tabular}%
}
\label{tab:prisonersdilemma}
\end{table}

\begin{table*}[h]

\caption[LLM Pythagorean Preference Relation Possible Outcomes]{Preference relation using positive and negative evaluation for preference and anti-preference.}
\centering
\resizebox{0.95\textwidth}{!}{%
\begin{tabular}{r|ll|llll|lll|}
\cline{2-10}
\multicolumn{1}{l|}{}                  & \multicolumn{2}{c|}{Strong Preference} & \multicolumn{4}{c|}{Partial Preference}                       & \multicolumn{3}{c|}{Indifference}     \\ \hline
\multicolumn{1}{|r|}{Best Evaluation}  & $L \succ M$        & $M \succ L$       & $L \succ M$   & $M \succ L$   & $L \sim M$    & $L \sim M$    & $L \sim M$ & $L \succ M$ & $M \succ L$ \\ \cline{1-1}
\multicolumn{1}{|r|}{Worst Evaluation} & $L \succ M$        & $M \succ L$       & $L \sim M$    & $L \sim M$    & $M \succ L$   & $L \succ M$   & $L \sim M$ & $M \succ L$ & $L \succ M$ \\ \hline
\multicolumn{1}{|r|}{Result}           & $L \succ M$        & $M \succ L$       & $L \succeq M$ & $M \succeq L$ & $M \succeq L$ & $L \succeq M$ & $L \sim M$ & $L \sim M$  & $L \sim M$  \\ \hline
\end{tabular}%
}
\label{tab:preference}
\end{table*}

\begin{figure*}[h]
\centering
\begin{subfigure}[t]{.48\linewidth}
  \centering
  \includegraphics[width=\linewidth]{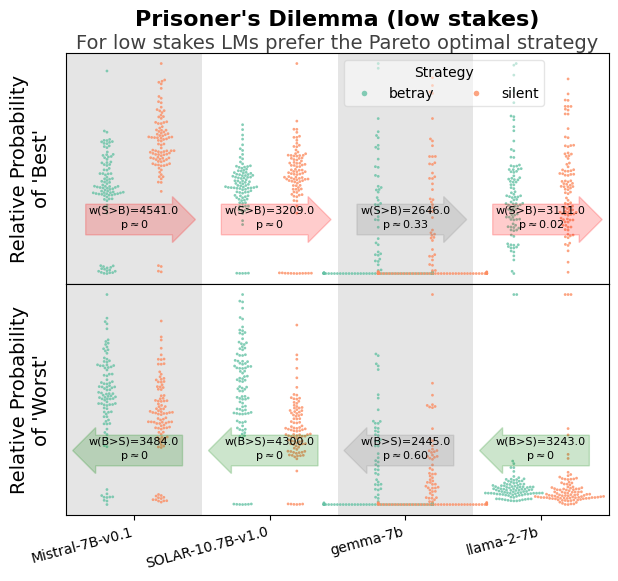}
\label{fig:PDLS}
\end{subfigure}%
\hfill
\begin{subfigure}[t]{.48\linewidth}
  \centering
  \includegraphics[width=\linewidth]{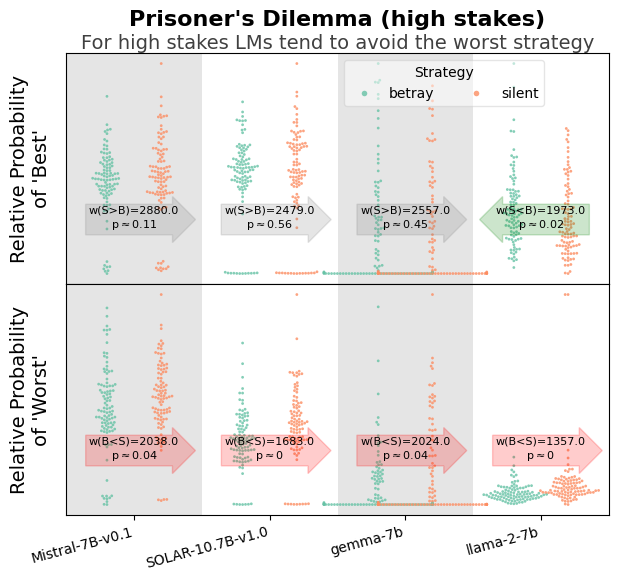}
  
\label{fig:PDHS}
\end{subfigure}
\caption[Summary of Prisoner's Dilemma Experiments]{Left: LLMs in a low stakes obfuscated prisoner's dilemma prefer cooperation. Right: LLMs in a high stakes obfuscated prisoner's dilemma prefer self-interest.}
\label{fig:PrisonerDilemma}
\end{figure*}

The Nash equilibrium strategy, defined as the option that obtains the best payoff without assuming that the opponent will change strategy \cite{nash1950non}, is typically expected to be chosen by rational agents in the PD (italics in Table \ref{tab:prisonersdilemma}). However, human behavior often deviates from this expectation. By choosing to remain silent, players can minimize the total number of months spent by either player in jail, known as the Pareto optimal strategy. 

\citet{yamagishi2016moral} conducted a large study on human subjects in Tokyo and showed that humans tend to cooperate (choose the Pareto optimal strategy) when the stake size is low. However, when the stake size is large, humans tend to choose to betray the other player in self-interest. The expected human behavior is highlighted in bold in Table \ref{tab:prisonersdilemma}.

\subsubsection{The Aim} Previous work found that in a non-repeated PD, GPT-4 tended to cooperate \cite{akata2023playing}. We extend this by considering open-source models, the effect of systematic variations, and the nuanced effect of stake size, which is known to change human behavior. To understand if LLMs tend to have human-like strategic preference for cooperation in the PD, we formulate RQ\ref{RQ:PD}:

\begin{theorem2}
When engaged in a PD, do LLMs tend to have cooperative preferences consistent with human preference including sensitivity to stake size?
\label{RQ:PD}
\end{theorem2}

\subsection{Methodology}

We utilize two scenarios to measure the impact of stake size: a low stakes version involving a shared air conditioner and a high stakes version involving a life support system.

In the low stakes scenario (prompt below), two individuals must decide whether to use (betray) or not use (silent) a shared air conditioner at night. The payoff matrix for this scenario is shown in the left side of Table \ref{tab:prisonersdilemma}. 

The high stakes scenario (prompt in appendix) follows a similar structure, with the shared air conditioner replaced by a life support system. The payoff matrix for this scenario, shown on the right side of Table \ref{tab:prisonersdilemma}.

To evaluate model preferences, we construct prompts that enumerate the options and possible results for each scenario. We then use counterfactual prompting to find the probability assigned to a constant evaluation word, as done in the previous experiment.

To control for alternative hypotheses, we construct permuted label versions of the prompt, with all permutations of label order and strategy assignment represented. We perform the experiments using both positive (``Best'') and negative (``Worst'') sentiment evaluation words. Finally, we conduct the experiments using populations (N=50) of Solar, Mistral, Gemma, and Llama-2 based on Table \ref{tab:VBP}.

\subsubsection{Low Stakes Prompt Example} My roommate and I share an AC. If one person uses the AC then they will get all and the other person will get none. If neither uses the AC then both will have a little. I do not know if my roommate will use the AC today, but they might. I cannot control their choice. Option1 is to use the AC. Option2 is to not use the AC. If I choose Option1 and my roommate uses the AC then I will be warm. If I choose Option1 and my roommate does not use the AC then I will be cold. If I choose Option2 and my roommate uses the AC then I will be hot. If I choose Option2 and my roommate does not use the AC then I will be cool. I want to be cold. I do not want to be hot or warm. My roommate wants to be cold and does not want to be hot or warm. Option1 is the \blackout{best}.

\subsubsection{Pythagorean Preference Relation} Inspired by work in pythagorean fuzzy preference relations for group decision making \cite{mandal2019pythagorean}, we consider that preference and anti-preference may vary independently. By measuring the probability of both positive and negative evaluation words, we arrive at measures related to the preference and anti-preference respectively.

To construct a preference relation from an intra-species population of LLMs, we use stratified population members generated with PopulationLM to evaluate the possible strategies. The result is a paired set of responses that permits the use of the non-parametric Wilcoxon rank-sum test. The null hypothesis for this test is that the distribution of observations of a single group, arising from two treatments, is not statistically different. Performing separate Wilcoxon tests on the positive and negative evaluations independently yields a measure and significance of both the preference and anti-preference over the strategies (betray and silent).

For strategies $L$ and $M$, each presented as options in context $c$, and a positive evaluation word used as the measure:

\begin{itemize}
\item If $p(e_{pos}|c,L) > p(e_{pos}|c,M)$ tends to hold in a population, as characterized by a Wilcoxon test, then we say the population has a significant preference for $L$ over $M$, denoted as $L \succ M$.

\item If $p(e_{pos}|c,L) > p(e_{pos}|c,M)$ tends to hold in a population, then we say the population has a significant anti-preference for $L$ over $M$, denoted as $M \succ L$.

\item If the result of a Wilcoxon test fails to be significant, then we say that the population has indifferent preference or anti-preference to $L$ over $M$, denoted $L \sim M$.

\end{itemize}

Table \ref{tab:preference} summarizes these possible resulting preferences based on the outcomes of the Wilcoxon tests for positive and negative evaluation words.

This preference relation is not transitive as it utilizes the Wilcoxon test \cite{lumley2016characterising}. However, transitive distribution tests may be counter productive as they are always reducible to univariate summary statistics \cite{lumley2016characterising}, and human preferences often fail to be transitive \cite{alos2022identifying}.

\subsection{Results: LLMs Prefer Cooperation in the PD}

In Figure \ref{fig:PrisonerDilemma} the probability of positive evaluation is shown in the top row and the probability of negative evaluation is shown in the bottom for all population members and all species. When the stakes are low, Solar, Mistral, and Llama-2 have a strong preference to cooperate. On the other hand, when the stakes are high, all models have a partial preference for self-interest. Interestingly, the Gemma population is uncertain regarding preference and anti-preference when faced with a low-stakes PD. This is most likely due to the brittleness result already discussed. 

In the high stakes scenario, Solar and Mistral show an anti-preference to cooperate (silent), but they don't prefer to act in self interest (betray). A human, choosing to use a life support system and potentially shorten the life of another, or choosing to trust another not to do the same, may ultimately experience a similar preference/anti-preference dichotomy. It's not preferable to potentially shorten the life of another. However, choosing to trust another individual to not act in self-preservation may be unacceptable. 

Addressing RQ\ref{RQ:PD}, the results indicate that self-consistent, non-brittle LLMs with VBP tend to have distinctly human-like cooperative preferences in the PD, including sensitivity to stake size. This holds true even when the scenario does not strongly resemble the classical incarnation of the dilemma.

\section{Do LLMs Have Human-Like Preference in the Traveler's Dilemma?}

The traveler's dilemma (TD), introduced by \cite{Basu1994}, is an interesting game that highlights scenarios in which humans tend to deviate from the Nash equilibrium.

\subsubsection{The Game} In the TD, two strangers traveling back from vacation have purchased the same antique, which the airline breaks. The individuals are informed independently and asked to provide the value of the antique within the range $[2,100]$. They are warned that overbidding the other passenger will result in a penalty.

Specifically, player A provides quote $Q_A$, and player B provides $Q_B$. The payoffs are determined as follows:

\begin{itemize}
\item If $Q_A > Q_B$, then the payoff for player A is $Q_B - 2$, and the payoff for player B is $Q_B + 2$.

\item If $Q_A < Q_B$, then the payoff for player A is $Q_A + 2$, and the payoff for player B is $Q_B - 2$.

\item If $Q_A = Q_B$, they receive the quoted value.

\end{itemize}

A strategy $Q_A$ is said to weakly dominate $Q_B$ if $Q_A$ is at least as good as $Q_B$ in all cases and provides a better payoff in at least one case \cite{Muthoo1996}. In the TD, quoting 99 weakly dominates quoting 100. Game theorists consider 100 to be eliminated as a strategy as 99 \textit{should be} strictly preferred. This creates a cascading elimination: iff 100 is removed, 98 weakly dominates 99.

The elimination of weakly dominated strategies results in a canonical Nash equilibrium that predicts rational players will quote the airline 2 dollars.

\begin{figure*}[t]
\centering

\begin{subfigure}[t]{.5\linewidth}
  \centering
  \includegraphics[width=\linewidth]{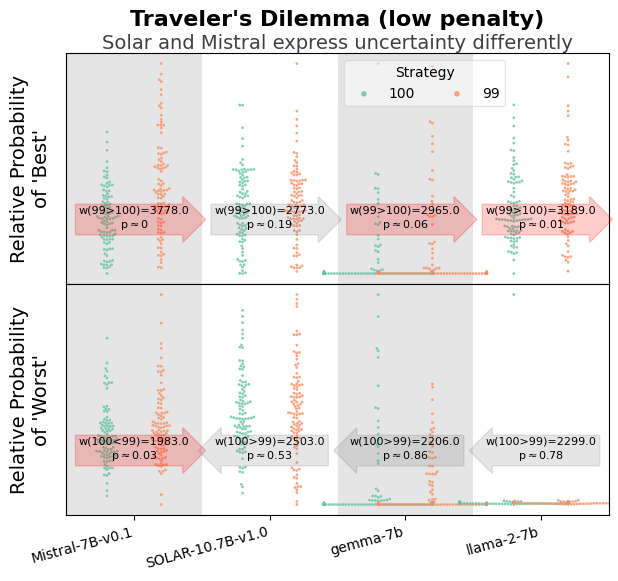}

\end{subfigure}%
\begin{subfigure}[t]{.5\linewidth}
  \centering
  \includegraphics[width=\linewidth]{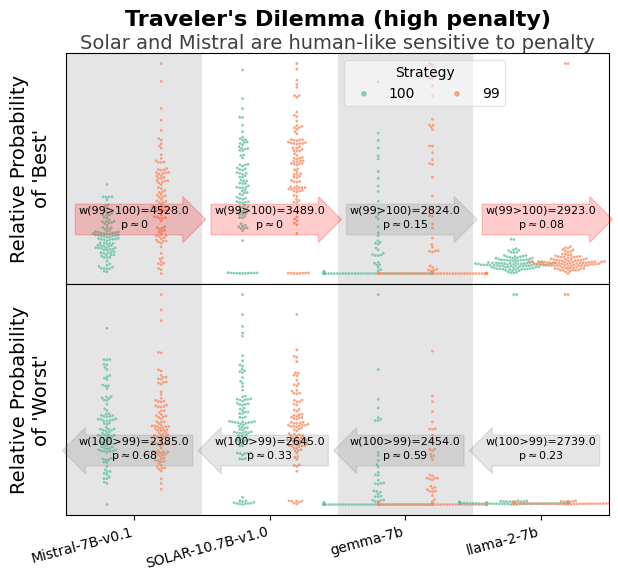}

\end{subfigure}

\caption[Summary of Traveler's Dilemma Experiments]{Left: LLM preference in a low penalty TD. Right: LLM preference in a high penalty TD}
\label{fig:TravelerDilemma}
\end{figure*}

\subsection{Human Behavior in the Traveler's Dilemma}

Empirical studies show that humans tend to prefer more cooperative strategies \cite{Becker2004}, choosing strategies in the mid-90s. However, when the penalty is increased, humans tend to choose strategies closer to the Nash equilibrium \cite{morone2014individual}, even though the penalty size has no game-theoretic effect on the equilibrium.

\citet{roberts2021finding} argues that human deviation from the Nash equilibrium suggests that humans are uncertain about their preference for 99 over 100, preventing the elimination of that strategy. They show that if this is the case, and the elimination scheme is retooled to permit fuzzy elimination, then human behavior is well predicted by fuzzy elimination of weakly dominated strategies. This explanation additionally captures the penalty size effect on the preference. 

\subsubsection{The Aim} Human deviation from the Nash equilibrium in the Traveler's Dilemma (TD) suggests that humans are indifferent toward strategies that weakly dominate more cooperative strategies when the penalty magnitude is small. This experiment investigates whether LLMs exhibit a similar penalty-based indifference toward dominated cooperative strategies. We examine the behavior of self-consistent LLMs with value-based preferences (VBP) in the TD by evaluating their preference and anti-preference for the strategies of quoting 99 and 100. Specifically, we formulate RQ\ref{RQ:TD}.

\begin{theorem2}
When engaged in a TD, do LLMs tend to prefer strategies closer the Nash equilibrium in response to increased penalty?
\label{RQ:TD}
\end{theorem2}

\subsection{Methodology}

To investigate LLM preferences in the Traveler's Dilemma (TD), we employ model species populations (N=50), counterfactual prompting, and the preference relation described in Table \ref{tab:preference}. The TD scenario, range of options, and payoffs for quoting 99 and 100 are provided in the prompt context. To control for superficial preference heuristics, we permute the labeling of options. All prompt patterns used in the experiments are in the technical appendix for transparency and reproducibility. We conduct two sets of experiments with penalty sizes of 2 and 20 to understand the effect of penalty size on the preference for cooperation.

\subsection{Results: LLMs Prefer Cooperation in the TD}

Figure \ref{fig:TravelerDilemma} shows the results for the high and low penalty scenarios. In the low penalty scenario, Solar and Mistral are indifferent to 99 and 100, that is $99 \sim 100$. However, when the penalty size increases to 20, Solar and Misral show a partial preference for 99, $99 \succeq 100$. 

Addressing RQ \ref{RQ:TD}, we find that non-brittle LLMs with VBP tend to have human-like preference for cooperation in the TD, including sensitivity to penalty size. LLMs with non-brittle VBP do not prefer 99 over 100 even though 100 is weakly dominated. Indifference to low-penalty, weakly dominated strategies prevents the elimination that leads to the canonical Nash equilibrium. Given this behavior was acquired from human data, it suggests this may hold for humans as well as previously proposed \cite{roberts2021finding}.


%
%
%

\section{Conclusion}

In this paper, we evaluated LLMs to identify candidates (SOLAR and Mistral) potentially useful for HRI tasks based on their human-like preference for cooperation. We found that value-based preference (VBP) and self-consistency tend to emerge as a function of model size and training token count but these are insufficient for reducing brittleness. We hypothesize that sliding window attention (SWA) may encourage more distributed representations and mitigate this. However, smaller models tend to prefer strategies based on superficial heuristics.

We showed that Solar and Mistral exhibit human-like cooperative preferences in both the Prisoner's Dilemma (PD) and Traveler's Dilemma (TD), including sensitivity to stake size and penalty size, respectively. These findings support the hypothesis for the origin of empirical deviation from the Nash equilibrium in the TD.

Our results contribute to understanding LLM cognitive behaviors and have implications for developing AI systems that better model human decision-making in strategic scenarios. Future research should focus on the impact of sliding window attention (SWA) on learned representations to develop more resilient and human-like language models for HRI applications.

\subsection{Limitations}



This work required a google colab based A100 GPU with 40GB of VRAM for approximately 5 hours to conduct the total set of experiments which yielded knowledge but necessarily contributed to resource consumption.

LLM based collaborators without appropriate safe guards pose a poorly understood risk that necessitates continued research and development.

%
%

\bibliography{aaai25}

\end{document}